\documentclass[a4paper,11pt]{article}
\usepackage{jheppub} 
\usepackage{lineno}

\title{Non-Invertible Duality Interfaces in Field Theories with Exotic Symmetries}

\author{Ryan C. Spieler}
\affiliation{Department of Physics and Weinberg Institute for Theoretical Physics, University of Texas at Austin\\
Austin TX 78712, USA}

\emailAdd{rcspieler.utexas.edu}

\abstract{In recent years, the concept of global symmetry has generalized considerably.  Two dramatic examples of this generalization are the exotic symmetries that govern theories with fractons and non-invertible symmetries, which do not fuse according to a group law.  Only recently has the interplay between these two been examined.  In this paper, we provide further examples of the interplay in the XY plaquette model, XY cube model, 1+1 d theory with global dipole symmetry, and the 2+1 d Lifshitz theory.  They are analogs of the duality symmetries in 2d CTFs and are constructed by first gauging a finite subgroup of the momentum symmetry on half of spacetime and then performing a duality transformation.  We analyze the fusion rules of the symmetries and find that they are condensation defects from an analog of higher gauging exotic symmetries.  We also address their dependence on the UV cutoff when relevant.}

\begin{document}
\maketitle
\flushbottom

\section{Introduction}
\subsection{(Non-Invertible) Symmetries}
The previous decade has seen an explosion of interest in quantum field theory's kinematic sector.  The insight guiding this, made clear in \cite{HighFSym} and reviewed in \cite{SnowmassGenSym,GenSymCondMat,NamekiRev,ShaoRev,LBRev}, is that symmetries in quantum field theory are topological defects \footnote{Throughout this paper, we use the term defect as a stand in for path integral insertions in a relativistic setting.  In non-relativistic settings, we are careful to distinguish defects (insertions extended in time) from operators (insertions that are not).} that act on operators by linking with them.  This includes not only familiar symmetries that surround points, but also ``higher form" symmetries that only act on extended operators.  These symmetries act like normal symmetries, they represent a group, they can lead to selection rules and conserved charges, they can be gauged, and they can have anomalies.  Moreover, different symmetries of different ``forms" can mingle, creating intricate higher group structures \cite{2Group1,2Group2}.

One dramatic generalization come from taking seriously the idea that topological defects are symmetries is that symmetries need not even form a group \cite{FiniteGauge,FiniteSubgroup,Line2d} but a more general structure \footnote{Thought to be a fusion higher category for finite symmetries.  One often hears the term ``symmetry category."} in which defects cannot be inverted.  Despite this oddity, there are several lines of thought arguing that these are symmetries.  They can be gauged or can have anomalies \cite{FiniteGauge,Line2d,FCS1,FCS2,2dAnomaly,BndyAnom,4danom1,4danom2}, imposing constraints on RG flows.  They have also been argued to be absent from quantum theories of gravity \cite{NICompleteness}.  There are several ways to construct non-invertible symmetries:
\begin{itemize}
    \item One can gauge a symmetry on a codimension $p>0$ submanifold of spacetime.  This is called higher gauging \cite{HigherGauging}.
    \item In theories with an ABJ anomaly, one can ``fix" the defect by introducing a Fractional Quantum Hall state on the defect, making the charge density closed and gauge invariant \cite{NIChiral,SM,NITR}.
    \item In theories with self-duality, one can gauge a finite subgroup of a symmetry on half of spacetime and invoke the duality to argue that one obtains a defect in a theory rather than an interface between two theories.  The defect, since it is constructed with topological boundary conditions, is topological.  This ``half gauging" construction is responsible for Kramers-Wannier type defects in various dimensions \cite{NonInvDual,NonInvKW,DualityTrialityETC,FreeTheory}.  \cite{GaugeNI,GaugeNI2} recently extended this construction to half gauging non-invertible symmetries.  We will pursue a different extension in this work.    
\end{itemize}
  We will demonstrate that the construction in the third bullet also applies to theories with exotic symmetries.  Similar symmetries appeared on the lattice in \cite{NISubSys}.
\subsection{Exotic Symmetries in Non-Relativistic Quantum Field Theory}

In recent years, peculiar symmetries that appear in non-relativistic theories have attracted considerable attention.  These symmetries, which  we will collectively refer to as exotic symmetries, owe their present popularity to the fact that they are the kinematic data of theories with fractons \cite{HermNandRev,PretkoRev,RadGromRev}, immobile excitations of exotic systems.   Fractons and other particles with restricted or absent mobility, and the strange symmetries that come with them, appear in a variety of field theories - both gapped and gapless \cite{XCubeQFT,Multipole,SS0,SS1,SS2,SS3,FCC,MoreExotic,TwistSym,TimelikeSymmetry}.  Those we explore in this paper broadly fall into two classes.  The first are subsystem symmetries those sensitive to the foliation \footnote{By foliation of a manifold $\mathcal{M}$, we mean a decompostion of $\mathcal{M}$ into the union of submanifolds called leaves.  Down to earth expositions about this can be found in \cite{Foliations,SAW,FoliatedQFT,ElectricMagneticModels,FoliatedExotic,ExoticFoliated}.} of the spacetime on which they live \cite{Foliations,SAW,FoliatedQFT,ElectricMagneticModels,FoliatedExotic,ExoticFoliated,FolMultipole}.  Here, the subsystem symmetries are symmetries on the leaves of the foliation.  This has many intriguing consequences, chief among them a number of charges subextensive in the lattice cutoff introduced to the theory.  The second class are global dipole symmetries \cite{TimelikeSymmetry,Graph,CptLifshitz}, which introduce an even more detailed dependence on UV data.  Both types of exotic symmetries appear in theories analogous to the $c=1$ compact boson - that is to say field theories with a self duality that we can use to construct non-invertible interfaces.  We do just that in this paper.  We note that some analogous interfaces are constructed on the lattice model for a different theory \footnote{In particular, \cite{NISubSys} considers an exotic version of the Ising model.  Our interfaces are constructed in exotic versions of the XY model, as well as in Lifshitz theories.} in \cite{NISubSys}.  Some related discussion on the lattice appears in \cite{KTss}.
\subsection{Outline of this Paper}
The remainder of this paper is structured as follows.  In section 2, we recapitulate the noninvertible duality symmetry in the $c=1$ compact boson, which may be viewed as the inspiration for this paper.  In section 3, we analyze the continuum limit of the $XY$ plaquette model.  We begin by reviewing the continuum theory introduced in \cite{SS1}, focusing on the global symmetries.  We then construct the noninvertible duality symmetry and analyze its fusion rules from a Lagrangian viewpoint, stressing the difference between a duality operator and duality defect.  We then analyze the situation on the lattice using the modified Villain model introduced in \cite{Villain}.  Section 4 repeats this analysis for the continuum limit of the XY cube theory.    

In section 5, we turn to the 1+1 dimensional theory with global dipole symmetry.  The continuum theory, discussed in \cite{TimelikeSymmetry} does not exhibit self duality, so we content ourselves with a treatment using a modified Villain model.  We review the theory, again emphasizing global symmetries, and construct the symmetry on the lattice, again analyzing its fusion with other operators and defects and clarifying the differences between symmetry operators and symmetry defects.  We repeat this analysis in section 6, this time analyzing the modified Villain formulation of the 2+1 dimensional Laplacian Lifshitz theory.

Section 7 summarizes the work and discusses multiple avenues for further progress.  The appendix provides more details on the deformation of defects.  
\section{Review of the Duality Interface in the $c=1$ Compact Boson}
In this appendix, we discuss the example that inspired this work - the duality interface in the $c=1$ compact boson theory.  Much of this material can be found in \cite{NonInvDual,FreeTheory,2dWeb}, and is included here for completeness.  The first presentation of the theory is based on a compact scalar subject to the identification $\phi \sim \phi + 2\pi$.  Its action is
\begin{equation}
    S = \frac{R^2}{4\pi}\int (d\phi)^2 .
\end{equation}
Due to the identification $\phi$ is not a well defined operator but $\exp[i\phi]$ and $\partial_\mu \phi$ are.  Just like the examples we consider, the theory has a second, dual presentation:
\begin{equation}
    S = \frac{\Tilde{R}^2}{4\pi}\int (d\Tilde{\phi})^2,
\end{equation}
where 
\begin{equation}
    \Tilde{R} = \frac{1}{R}.
\end{equation}
The duality in question is T-duality (a 2d version of electric-magnetic duality), expressed as
\begin{equation}
    -iR^2 d\phi = \star d \Tilde{\phi}.
\end{equation}

Let us motivate the interface.  We will gauge the $\mathbb{Z}_N$ subgroup of the $U(1)$ momentum symmetry \footnote{By momentum, we mean target space momentum.  This terminology comes from String Theory, in which the target space is spacetime so there is no risk of confusion.  We will use this terminology throughout the paper.  The reader might also hear this called shift or particle number symmetry.}.  This shrinks the range of $\phi$ so that $\phi \in [0,\frac{2\pi}{N})$.  To get a variable with the same range as the original, define $\hat{\phi} = N\phi \in [0,2\pi)$ whose action is
\begin{equation}
    S = \frac{R^2}{4\pi N^2}\int (d\hat{\phi})^2 .
\end{equation}
We see that we have shrunk $R$ to $R/N$.  Note that if we set $R = \sqrt{N}$, this maps $R \rightarrow \frac{1}{R}$, which is undone by T-duality!  Thus, we construct the interface at $R = \sqrt{N}$.  To gauge the $\mathbb{Z}_N$ subgroup of the $U(1)$ momentum symmetry, we add a gauge field $a \sim a + d\alpha$ in the usual way, giving the action
\begin{equation}
    S = \int[\frac{N}{4\pi}(d\phi -a)^2 + \frac{iN}{2\pi}\sigma da].
\end{equation}
The first term makes the momentum symmetry a redundancy.  The second term imposes 
\begin{equation}
    [\frac{a}{2\pi}] \in H^1(M;\mathbb{Z}_N)
\end{equation}
i.e. it forces $a$ to be a $\mathbb{Z}_N$ gauge field.  It is known that when one gauges a $\mathbb{Z}_N$ zero-form symmetry, the resulting theory has a dual (or quantum) $\mathbb{Z}_N$ $d-2 (=0)$ form symmetry generated by the background field for the original symmetry.  Thus, the gauged theory has the operator:
\begin{equation}
    \eta = \exp[i\oint a] ; \eta^N = 1.
\end{equation}
To construct the defect, we will consider a codimension one submanifold $S$ of the spacetime $M$ and impose $a\vert_S = 0$.  Since $da=0$, this is a topological boundary condition \footnote{In more detail, consider deforming the boundary while preserving the boundary condition. This amounts to inserting a chunk of gauge theory on a contractible two-manifold with Dirichlet boundary conditions on $a$.  Since $a$ is flat, all observables on such a manifold are trivial, and inserting the chunk will not affect the path integral.  We thank Po-Shen Hsin for clarifying this.}.  We call the line with boundary condition $\mathcal{D}(S)$.  Since $\mathcal{D}(S)$ is defined as a line that kills $a$, we immediately obtain the fusion rule:
\begin{equation}
    \mathcal{D} \times \eta = \eta \times \mathcal{D} = \mathcal{D}.
\end{equation}
To determine the fusion rule of $\mathcal{D}$, we introduce a Lagrangian presentation of the interface.  It is:
\begin{equation}
    S = \frac{N}{4\pi}\int_L (d\phi_L)^2 + \frac{N}{4\pi}\int_R (d\phi_R)^2 + \frac{iN}{2\pi}\int_S \phi_L d\phi_R,
\end{equation}
where $L$ ($R$) is the region to the left (right) of the interface.  $\phi_L$ ($\phi_R$) is the field on $L$ ($R$).  To preserve the variational principle in the presence of the interface, we need to demand:
\begin{equation}
    d\phi_L\vert_S = i\star d\phi_R \vert_S  .
\end{equation}
Applying T-duality lets us write 
\begin{equation}
    d\phi_L\vert_S = \frac{1}{N}d\Tilde{\phi}_R \vert_S =  i\star d\phi_R \vert_S .
\end{equation}
Therefore, this interface does exactly what we want - it gauges $\mathbb{Z}_N$ on $R$ and applies T-duality.  Now that we know our interface behaves as desired, let us compute the fusion rules.  Consider two interfaces on $S$ and $S'$.  Our action is
\begin{equation}
    S = \frac{N}{4\pi}\int_L (d\phi_L)^2 + \frac{N}{4\pi}\int_I (d\phi_I)^2 + \frac{N}{4\pi}\int_R (d\phi_R)^2 + \frac{iN}{2\pi}\int_S \phi_L d\phi_I + \frac{iN}{2\pi}\int_{S'} \phi_I d \phi_R ,
\end{equation}
where $I$ is the region between $S$ and $S'$ and $\phi_I$ is the field on $I$.  When $S=S'$, this reduces to
\begin{equation}
    S = \frac{N}{4\pi}\int_L (d\phi_L)^2 + \frac{N}{4\pi}\int_R (d\phi_R)^2 + \frac{iN}{2\pi}\int \phi_I d(\phi_R-\phi_L).
\end{equation}
To understand the fusion rule, note that integrating out $\phi_I$ imposes
\begin{equation}
    [\frac{\phi_R-\phi_L}{2\pi}] \in H^0(S;\mathbb{Z}_N)
\end{equation}
i.e. $\phi$ gets shifted by a $\frac{2\pi}{N}\mathbb{Z}$ value across $S$.  This is nothing but the action of the $\mathbb{Z}_N$ subgroup of the momentum symmetry!  Thus, we have discovered the non-invertible fusion rule:
\begin{equation}
    \mathcal{D} \times \mathcal{D} = \sum_{a = 1}^N \eta^a .
\end{equation}
The fusion rules we discussed through the section are the fusion rules of a Tambara-Yamagami fusion category for $\mathbb{Z}_N$.  For $N=2$ it reduces to the fusion rules for the Kramer-Wannier defect.  For $N=1$ it becomes invertible and describes the fusion rules for a T-duality defect.  Let's make a couple comments about the fusion rule:
\begin{itemize}
    \item One can deduce the fusion rule up to normalization by demanding that it be consistent with the rule $\mathcal{D} \times \eta$.
    \item The sum of possible lines along the interface is an example of a condensation defect \cite{HigherGauging}.  The proliferation of symmetry lines along the interaface is interpreted as gauging the symmetry on that interface.
\end{itemize}
\section{Duality Symmetry in the XY Plaquette Model}
In this section, we construct and analyze the noninvertible duality symmetry in the continuum limit of the XY Plaquette model.  We will also derive the duality symmetry on the lattice, using a modified Villain model for the XY Plaquette model.  
\subsection{Review of the Theory}
We begin by reviewing the theory.  Many more details can be found \cite{SS1}.  One continuum presentation of the XY Plaquette Model is in terms of a scalar $\phi$ subject to the identification $\phi\sim\phi + 2\pi n^x(x) + 2\pi n^y(y)$, where $n^i(x^i)$ are integer valued functions of the appropriate coordinate.  Its Lagrangian is
\begin{equation}
    \mathcal{L} = \frac{\mu_0}{2}(\partial_\tau \phi)^2 + \frac{1}{2\mu}(\partial_x \partial_y \phi)^2.
\end{equation}
Thanks to the identification on $\phi$, operators such as $\phi$ and $\partial_i  \phi$ are not well defined, but operators such as $\exp[i\phi]$ and $\partial_x \partial_y \phi$ are.  The theory has two symmetries that will play a key role in the following.  The first is a $U(1)$ momentum  dipole symmetry that shifts $\phi$.  It follows from the equation of motion:
\begin{equation}
    \partial_\tau J^\tau = \partial_x \partial_y J^{xy},
\end{equation}
where
\begin{equation}
    J^\tau = i\mu_0 \partial^\tau \phi ; J^{xy} = i\frac{1}{\mu}\partial^x \partial^y \phi .
\end{equation}
The symmetry operators for the momentum dipole symmetry are 
\begin{equation}
    Q^x_m(x) = \oint dy J^\tau ; Q^y_m(y) = \oint dx J^\tau.
\end{equation}
They satisfy the constraint
\begin{equation}
    \oint dx Q^x_m(x) = \oint dy Q^y_m(y).
\end{equation}
Thus, on a square lattice with $L_x \times L_y$ sites, there are $L_x+L_y-1$ operators.
The theory also has a $U(1)$ winding \footnote{The author has also heard this called vorticity.} dipole symmetry with currents
\begin{equation}
    J_\tau^{xy} = \frac{1}{2\pi}\partial^x\partial^y \phi ; J = \frac{1}{2\pi} \partial^\tau \phi,
\end{equation}
which obey the obvious continuity equation:
\begin{equation}
    \partial_\tau J_\tau^{xy} = \partial^x \partial^y J.
\end{equation}
The symmetry operators are
\begin{equation}
    Q^x_w(x) = \oint dy J_\tau^{xy} ; Q_w^y(y) = \oint dx J_\tau^{xy}.
\end{equation}
They satisfy the constraint
\begin{equation}
    \oint dx Q^x_w(x) = \oint dy Q^y_w(y).
\end{equation}
Thus, on a square lattice with $L_x \times L_y$ sites, there are $L_x+L_y-1$ operators.
The theory has additional symmetries if one places it on a peculiar manifold such as the twisted torus in \cite{TwistSym}, but we do not consider such cases in this paper.

A second, dual, presentation of the theory is in terms of a compact field $\phi^{xy}$ subject to the identification $\phi^{xy} \sim \phi^{xy} + 2\pi(n^x(x) + n^y(y))$, where $n^i(x^i)$ are as above.  Its Lagrangian is
\begin{equation}
    \mathcal{L} = \frac{\Tilde{\mu}_0}{2}(\partial_\tau \phi^{xy})^2 + \frac{1}{2\Tilde{\mu}}(\partial_x \partial_y \phi^{xy})^2.
\end{equation}
Identical comments about which operators are well defined apply here.  If one explicitly dualizes the $\phi$ presentation of the theory, one can show that the tilded parameters are related to their untilded counterparts as 
\begin{equation}
    \Tilde{\mu}_0 = \frac{\mu}{4\pi^2}; \Tilde{\mu} = 4\pi^2 \mu_0.
\end{equation}
The theory has two symmetries will discuss.  It has a dual $U(1)$ momentum dipole symmetry.  It follows from the equation of motion:
\begin{equation}
    \partial_\tau J_\tau^{xy} = \partial^x \partial^y J,
\end{equation}
where
\begin{equation}
    J_\tau^{xy} = i\Tilde{\mu}_0 \partial_\tau \phi^{xy} ; J = i\frac{1}{\Tilde{\mu}}\partial_x\partial_y \phi^{xy}.
\end{equation}
The theory also has a dual $U(1)$ winding dipole symmetry with currents
\begin{equation}
    J_\tau = \frac{1}{2\pi}\partial_x\partial_y \phi^{xy}; J^{xy} =\frac{1}{2\pi}\partial_\tau \phi^{xy} 
\end{equation}
satisfying the obvious continuity equation:
\begin{equation}
    \partial_\tau J_\tau^{xy} = \partial_x \partial_y J.
\end{equation}
Analogizing the currents that furnish the same representation of $\mathbb{Z}_4$ gives the duality relation:
\begin{equation}
   \partial_x \partial_y \phi = i\frac{\mu}{2\pi}\partial_\tau \phi^{xy}; \partial_\tau \phi = \frac{i}{2\pi \mu_0}\partial_x \partial_y \phi^{xy}.
\end{equation}
Thus, we see that this duality is similar to the well known T-duality of the $c=1$ compact boson, discussed in the previous section.  For that reason, we call this duality, and analogous dualities that will appear throughout this paper T-duality.

\subsection{Constructing the Symmetry}
We now move to construct the symmetry from half gauging and examine its interaction with other operators and defects in the theory.  We construct the symmetry by gauging a $\mathbb{Z}_N$ subgroup of the $U(1)$ momentum dipole symmetry.  This restricts the range of $\phi$ so that we can write an identical looking Lagrangian by defining $\hat{\phi} = N\phi$:
\begin{equation}
    \mathcal{L} = \frac{\mu_0}{2N^2}(\partial_\tau \phi)^2 + \frac{1}{2\mu N^2}(\partial_x\partial_y \phi)^2
\end{equation}
Thus, we see that the effect of gauging is to map $\mu_0$ to $\frac{\mu_0}{N^2}$ and $\mu$ to $\mu N^2$.  Remarkably, T-duality undoes this for any self dual values of $\mu$ and $\mu_0$.  Let us now gauge $\mathbb{Z}_N$ in detail.  The appropriate Lagrangian is:
\begin{equation}
    \mathcal{L} = \frac{\mu_0}{2}(\partial_\tau \phi - A_\tau)^2 + \frac{1}{2\mu}(\partial_x\partial_y\phi - A_{xy})^2 + \frac{iN}{2\pi}\phi^{xy}(\partial_\tau A_{xy} - \partial_x \partial_y A_\tau).
\end{equation}
Here, $A_\tau$ and $A_{xy}$ are $U(1)$ gauge fields that couple to $J^\tau$ and $J^{xy}$, respectively.  Thanks to the continuity equation, they have the redundancy:
\begin{equation}
    A_\tau \sim A_\tau + \partial_\tau \alpha; A_{xy} \sim A_{xy} + \partial_x \partial_y \alpha.
\end{equation}
Integrating out $\phi^{xy}$ imposes the constraint:
\begin{equation}
    \frac{N}{2\pi}(\partial_\tau A_{xy} - \partial_x \partial_y A_\tau) = 0
\end{equation}
which implies that the gauge fields are $\mathbb{Z}_N$ gauge fields.  Adding the term is the analog of adding a BF term to gauge a $\mathbb{Z}_N$ subgroup of a $U(1)$ symmetry.  The gauged theory has the symmetry defect:
\begin{equation}
    \eta_\tau(x,y) = \exp[i\oint d\tau A_\tau] ; \eta_\tau(x,y)^N = 1,
\end{equation}
with the second equality following from the equation of motion.  It also has the symmetry operator
\begin{multline}
    \eta_{xy}(x_1,x_2) = \exp[i\int_{x_1}^{x_2} dx\oint  dy A_{xy}];\eta_{xy}(y_1,y_2) = \exp[i\int_{y_1}^{y_2} dy\oint  dx A_{xy}];\\ \eta_{xy}(x_1,x_2)^N = \eta_{xy}(y_1,y_2)^N = 1,
\end{multline}
with the equalities in the second line following from the equation of motion.  We interpret these as dual symmetries to the momentum dipole symmetry.  To construct the noninvertible symmetry, we impose Dirichlet boundary conditions on $(A_\tau,A_{xy})$ along a codimension 1 surface $\mathcal{S}$.  We call the symmetry constructed in this way $\mathcal{D}(\mathcal{S})$.   There are two basic possibilities:
\begin{itemize}
    \item $\mathcal{S}$ exists on a time slice.  In this case, $\mathcal{D}(\mathcal{S})$ is an operator on the Hilbert space of the theory.
    \item $\mathcal{S}$ extends in the time direction.  In this case, $\mathcal{D}(\mathcal{S})$ is a defect.  It changes the Hilbert space relative to the case without it. 
\end{itemize}
Since we are working with a non-relativistic theory, these two possibilities are fundamentally distinct, and we will treat them separately.   
\subsection{Symmetry Operator}

For simplicity, we consider an operator along $\tau=0$.  The action of the system with the operator is 
\begin{multline}
    S = \int_{\tau\leq 0}d\tau dx dy [\frac{\mu_0 }{2}(\partial_\tau \phi_-)^2 + \frac{1}{2\mu}(\partial_x\partial_y\phi_-)^2] - \frac{iN}{2\pi}\int_{\tau =0}dx dy[\phi_- \partial_x \partial_y \phi_+] \\ + \int_{\tau \geq 0} d\tau dx dy [\frac{\mu_0}{2}(\partial_\tau \phi_+)^2 + \frac{1}{2\mu}(\partial_x\partial_y\phi_+)^2]
\end{multline}
Why is this the symmetry operator?  In order to preserve the variational principle that is present without the operator, we need to demand 
\begin{equation}
    \mu_0\partial_\tau \phi_- = -\frac{iN}{2\pi}\partial_x \partial_y \phi_+
\end{equation}
at $\tau=0$.  Applying the T-duality formula gives 
\begin{equation}
    \frac{1}{N}\partial_\tau \phi_- = \partial_\tau \phi^{xy}_+= -\frac{2\pi i}{\mu}\partial_x \partial_y \phi_+.
\end{equation}
We interpret the first equality as gauging $\mathbb{Z}_N$ and appropriately shrinking the scalar and the second as applying the duality.  Thus, this operator does what we want.  Let us see what happens when we fuse an operator and its orientation reversal $\Bar{\mathcal{D}}$.  The action for two operators is
\begin{multline}
    S = \int_{\tau\leq 0}d\tau dx dy [\frac{\mu_0 }{2}(\partial_\tau \phi_-)^2 + \frac{1}{2\mu}(\partial_x\partial_y\phi_-)^2] - \frac{iN}{2\pi}\int_{\tau =0}dx dy[\phi_- \partial_x \partial_y \phi_I] \\ + \int_{\epsilon \geq \tau \geq 0} d\tau dx dy [\frac{\mu_0 }{2}(\partial_\tau \phi_I)^2 + \frac{1}{2\mu}(\partial_x\partial_y\phi_I)^2] + \frac{iN}{2\pi}\int_{\tau = \epsilon} dx dy [\phi_I \partial_x \partial_y \phi_+] \\+ \int_{\tau\geq \epsilon} d\tau dx dy [\frac{\mu_0 }{2}(\partial_\tau \phi_+)^2 + \frac{1}{2\mu}(\partial_x\partial_y\phi_+)^2].
\end{multline}
Fusion corresponds to taking $\epsilon \rightarrow 0$, in which case action on the operator becomes
\begin{equation}
    S_{op} = \frac{iN}{2\pi}\int dx dy \phi_I \partial_x \partial_y(\phi_+ - \phi_-).
\end{equation}  
To interpret this fusion rule, note that integrating out $\phi_I$ imposes
\begin{equation}
    \partial_x \partial_y (\phi_+-\phi_-) =0 ; \phi_+-\phi_- \in \frac{2\pi}{N}\mathbb{Z}.
\end{equation}
  Thus, the product defect shifts the action of $\eta_{xy}$ to some power.  We have uncovered the fusion rule
\begin{equation}
    \mathcal{D}\times \Bar{\mathcal{D}} = \sum_{k}(\sum_{i=1}^N (\eta_{xy}(x_k,x_{k+1})^i + \eta_{xy}(y_k,y_{k+1})^i)).
\end{equation}
Note that this gauges the a $\mathbb{Z}_N$ subgroup of the $U(1)$ momentum dipole symmetry on the surface on which the operator is supported.  Thus, we can view the right hand side as a condensation defect obtained from higher gauging.  It requires a lattice regularization for the sum to be discrete.  This is the field theoretic version of the grid operator in \cite{NISubSys}.  By dragging $\exp[i\phi]$ across the operator to the gauged half of spacetime, it acquires the tail needed to be gauge invariant, so this symmetry implements the usual order-disorder transition.
\subsection{Symmetry Defect}

For simplicity, we consider a defect along $y=0$.  The action of the system with the defect is 
\begin{multline}
    S = \int_{y\leq 0}d\tau dx dy [\frac{\mu_0 }{2}(\partial_\tau \phi_L)^2 + \frac{1}{2\mu}(\partial_x\partial_y\phi_L)^2] + \frac{iN}{2\pi}\int_{y=0}d\tau dx[\phi_L \partial_\tau \partial_x \phi_U] \\ + \int_{y \geq 0} d\tau dx dy [\frac{\mu_0 }{2}(\partial_\tau \phi_U)^2 + \frac{1}{2\mu}(\partial_x\partial_y\phi_U)^2].
\end{multline}
To see that this is the desired defect, note that, in order to preserve the variational principle, we need to require
\begin{equation}
    \frac{iN}{2\pi} \partial_\tau \partial_x \phi_U = -\frac{1}{\mu} \partial_x^2 \partial_y \phi_L
\end{equation}
at $y=0$.  Applying the T-duality formula gives 
\begin{equation}
     \frac{1}{N}\partial_x^2 \partial_y \phi_L = \partial_x^2 \partial_y \phi_U^{xy} = -2\pi \mu_0 i  \partial_\tau \partial_x \phi_U .
\end{equation}
We interpret the first equality as gauging $\mathbb{Z}_N$ and appropriately shrinking the scalar and the second as applying the duality.  Therefore, this defect does what we want.  Let us see what happens when we fuse a defect and its orientation reversal.  The action for two defects is 
\begin{multline}
    S = \int_{y \leq 0}[\frac{\mu_0 }{2}(\partial_\tau \phi_L)^2 + \frac{1}{2\mu}(\partial_x\partial_y\phi_L)^2] - \frac{iN}{2\pi}\int_{y=0} d\tau dx [\phi_L \partial_\tau \partial_x \phi_I]\\ + \int_{\epsilon \geq y \geq 0}[\frac{\mu_0 }{2}(\partial_\tau \phi_I)^2 + \frac{1}{2\mu}(\partial_x\partial_y\phi_I)^2] + \frac{iN}{2\pi}\int_{\tau=0} d\tau dx [\phi_I \partial_\tau \partial_x \phi_U] \\ + \int_{y \geq \epsilon}[\frac{\mu_0 }{2}(\partial_\tau \phi_U)^2 + \frac{1}{2\mu}(\partial_x\partial_y\phi_U)^2]. 
\end{multline}
Fusion corresponds to taking $\epsilon \rightarrow 0$.  The resulting defect action is
\begin{equation}
    S = \frac{iN}{2\pi}\int_{y=0} d\tau dx [\phi_I\partial_\tau\partial_x(\phi_U - \phi_R)].
\end{equation}
To understand this, note that integrating out $\phi_I$ imposes
\begin{equation}
    \partial_\tau \partial_x(\phi_U-\phi_L) =0; \phi_U-\phi_L \in \frac{2\pi}{N}\mathbb{Z}.
\end{equation}
  Thus, the product defect is an action of $\eta_\tau$ to some power.  We have uncovered the fusion rule:
\begin{equation}
    \Bar{\mathcal{D}} \times \mathcal{D} = \sum_k \sum_{i=1}^N \eta_\tau (x_k)^i.
\end{equation}
Note that we require a lattice regularization along the x direction to make the above sum discrete.  The right hand side gauges a $\mathbb{Z}_N$ subgroup of the $U(1)$ momentum dipole symmetry along the defect.  Thus, we can view it as a condensation defect from higher gauging.  This is the field theoretic version of the grid defect in \cite{NISubSys}.  
\subsection{Modified Villain Construction}

In this subsection, we construct both the non-invertible operator and non-invertible defect on a Euclidean lattice.  This construction deduces the form of the operator and defect actions rather than asserting them.  It also allows a precise analysis of the deformation of the defect, as discussed in the appendix.   We follow the strategy in \cite{NonInvDual}.  More details on the lattice model we use can be found in \cite{Villain}.  The model lives on the Euclidean cubic lattice $\mathbb{Z}_{L_\tau} \times \mathbb{Z}_{L_x} \times \mathbb{Z}_{L_y}$.  The action takes the form:
\begin{multline}
    S = \frac{\mu_0}{2}\sum_{\tau-links} (\Delta_\tau \phi - 2\pi n_\tau)^2 + \frac{1}{2\mu}\sum_{xy-plaquettes}(\Delta_x \Delta_y \phi - 2\pi n_{xy})^2 \\+ i\sum_{cubes}\phi^{xy}(\Delta_\tau n_{xy} - \Delta_x\Delta_y n_\tau)
\end{multline}
Here, $\Delta_i$ is the lattice derivative along a link in the $i^{th}$ direction, $\phi$ is a real-valued variable on a site, $n_\tau$ is an integer-valued variable on a $\tau$ link, $n_{xy}$ is an integer-valued variable on a plaquette in the xy plane, and $\phi^{xy}$ is a real-valued variable on a cube (dual site).  The fields are subject to the following identifications:
\begin{equation}
    \phi \sim \phi + 2\pi k
\end{equation}
\begin{equation}
    n_\tau \sim n_\tau + \Delta_\tau k
\end{equation}
\begin{equation}
    n_{xy} \sim n_{xy} + \Delta_x \Delta_y k,
\end{equation}
\begin{equation}
    \phi^{xy} \sim \phi^{xy} + 2\pi \Tilde{k}^{xy}
\end{equation}
where $k$ and $\Tilde{k}^{xy}$ are integer-valued variables on (dual) sites.  The first and fourth identifications compactify the target space.  If we view $n$ as a gauge field, then $\phi^{xy}$ enforces a flatness constraint.  Dualizing theories such as this on the lattice is a matter of applying the Poisson resummation formula:
\begin{multline}
    \sum_n \exp[-\frac{\beta}{2}(\theta - 2\pi n)^2 + in\Tilde{\theta}] \\ = \sum_{\Tilde{n}}\exp[-\frac{1}{4\pi^2 \beta}(\Tilde{\theta} - 2\pi \Tilde{n})^2 -\frac{i\theta}{2\pi}(2\pi \Tilde{n} - \Tilde{\theta})].
\end{multline}
For instance, applying this for both $n_\tau$ and $n_{xy}$ above gives a dual presentation of the model, with action:
\begin{multline}
    S = \frac{\Tilde{\mu}_0}{2}\sum_{dual-\tau-links} (\Delta_\tau \phi^{xy} - 2\pi \Tilde{n}^{xy}_\tau)^2 + \frac{2}{\Tilde{\mu}}\sum_{dual-xy-plaqs}(\Delta_x \Delta_y \phi^{xy} - 2\pi \Tilde{n})^2 \\+ i\sum_{sites}\phi(\Delta_\tau \Tilde{n} - \Delta_x \Delta_y \Tilde{n}^{xy}_\tau ), 
\end{multline}
where
\begin{equation}
    \Tilde{n}^{xy}_\tau \sim  \Tilde{n}^{xy}_\tau + \Delta_\tau \Tilde{k}^{xy}
\end{equation}
\begin{equation}
    \Tilde{n} \sim \Tilde{n} + \Delta_x \Delta_y \Tilde{k}^{xy}.
\end{equation}
Everything in the above action has the same interpretation as the $\phi$ action, just on the dual lattice.  

We proceed as in the continuum, beginning by gauging the $\mathbb{Z}_N$ subgroup of the $U(1)$ momentum dipole symmetry.  The gauged action is
\begin{multline}
    S = \frac{\mu_0}{2}\sum_{\tau-links} (\Delta_\tau \phi - 2\pi n_\tau - \frac{2\pi}{N}\Hat{n}_\tau)^2 + \frac{1}{2\mu}\sum_{xy-plaquettes}(\Delta_x \Delta_y \phi - 2\pi n_{xy} - \frac{2\pi}{N}\hat{n}_{xy}) \\+ i\sum_{cubes}\phi^{xy}(\Delta_\tau n_{xy} - \Delta_x\Delta_y n_\tau + \frac{1}{N}\Delta_\tau \hat{n}_{xy} -\frac{1}{N}\Delta_x \Delta_y \hat{n}_\tau) \\+ \frac{2\pi i}{N}\sum_{cubes} \hat{m}^{xy}(\Delta_\tau \hat{n}_{xy} -\Delta_x \Delta_y \hat{n}_\tau).
\end{multline}
The fields are subject to the identifications:
\begin{equation}
    \phi \sim \phi + 2\pi k + \frac{2\pi q}{N}
\end{equation}
\begin{equation}
    n_\tau \sim n_\tau + \Delta_\tau k - l_\tau
\end{equation}
\begin{equation}
    n_{xy} \sim n_{xy} + \Delta_x \Delta_y k - l_{xy}
\end{equation}
\begin{equation}
    \phi^{xy} \sim \phi^{xy} + 2\pi \Tilde{k}^{xy}
\end{equation}
\begin{equation}
    \Hat{n}_\tau \sim \Hat{n}_\tau + \Delta_\tau q + Nl_\tau
\end{equation}
\begin{equation}
    \Hat{n}_{xy} \sim \Hat{n}_{xy} + \Delta_x \Delta_y q + N l_{xy}
\end{equation}
\begin{equation}
    \Hat{m}^{xy} \sim \Hat{m}^{xy} - \Tilde{k}^{xy} + N q^{xy}.
\end{equation}
Above, $q$ is an integer-valued variable on a site, $q^{xy}$ is an integer-valued variable on a cube, $l_\tau$ is an integer-valued variable on a $\tau$ link, $l_{xy}$ is an integer-valued on an $xy$ plaquette, $\hat{n}_\tau$ is an integer-valued variable on a $\tau$ link, $\hat{n}_{xy}$ is an integer-valued variable on an $xy$ plaquette, and $\hat{m}^{xy}$ is an integer-valued variable on a cube.   We can gauge away $\Hat{m}^{xy}$, $n_\tau$, and $n_{xy}$ and redefine variables: 
\begin{equation}
    \varphi := N \phi ; \varphi^{xy} := \frac{\phi^{xy}}{N}.
\end{equation}
This yields the action:
\begin{multline}
    S = \frac{\mu_0}{2N^2} \sum_{\tau-links} (\Delta_\tau \varphi - 2\pi \hat{n}_\tau)^2 + \frac{1}{2N^2 \mu}\sum_{xy-plaquettes}(\Delta_x \Delta_y \varphi - 2\pi \hat{n}_{xy})^2 \\ +i \sum_{cubes} \varphi^{xy}(\Delta_\tau \hat{n}_{xy} - \Delta_x \Delta_y \hat{n}_\tau).
\end{multline}
Note that gauging $\mathbb{Z}_N$ indeed rescales the fields as anticipated from our continuum analysis.

To construct the symmetry, we simply apply the above to the theory on half of spacetime.  Note that we acquire a boundary term in killing the $n$s.  This term is different for the operator and the defect, so we will discuss it on a case by case basis.  Let's start with what we can say independent of whether the symmetry is an operator or defect.  In particular, we can apply the Poisson resummation formula to the theory on the half of spacetime on which we gauge it.  The resulting action is
\begin{multline}
    S = \frac{\Tilde{\mu}_0}{2}\sum_{\tau-links} (\Delta_\tau \varphi^{xy} - 2\pi \Tilde{n}_\tau)^2 + \frac{1}{2\Tilde{\mu}}\sum_{xy-plaquettes}(\Delta_x \Delta_y \varphi^{xy} - 2\pi \Tilde{n}_{xy})^2 \\ + \frac{i}{2\pi}\sum_{cubes} \Delta_\tau \varphi (2\pi \Tilde{n}_{xy} - \Delta_x \Delta_y \varphi^{xy}) + \frac{i}{2\pi}\sum_{cubes} \Delta_x \Delta_y \varphi (2\pi \Tilde{n}_\tau - \Delta_\tau \varphi^{xy}) + S_\partial
\end{multline}
We can sum by parts to obtain the $\varphi^{xy}$ presentation of the theory.  Since this clearly depends on whether we have an operator or a defect, we will treat them in turn.  If the symmetry is an operator, the initial boundary term is
\begin{equation}
    S_\partial = -iN \sum \varphi^{xy}\hat{n}_{xy}.
\end{equation}
After summing by parts, we have
\begin{equation}
    S_{operator} = \frac{iN}{2\pi}\sum \phi (2\pi \Tilde{n}_{xy}-\Delta_x\Delta_y \varphi^{xy}) - iN\sum \varphi^{xy} \hat{n}_{xy}.
\end{equation}
In the continuum, this becomes the BF type term we asserted earlier.  Let's discuss the fusion between this operator and its opposite - namely the operator constructed by gauging the $\mathbb{Z}_N$ subgroup of the momentum dipole symmetry in the dual theory on the $\tau \leq 0$ side.  That operator is
\begin{equation}
    \Bar{S}_{operator} = \frac{iN}{2\pi}\sum \varphi^{xy}(\Delta_x \Delta_y \phi - 2\pi n_{xy}) + iN\sum \phi \hat{\Tilde{n}}_{xy}
\end{equation}
Placing the two on top of each other and gauging away $\Tilde{n}_{xy}$, $\hat{n}_{xy}$, $n_{xy}$, and $\hat{\Tilde{n}}_{xy}$ \footnote{Aside from residual gauge symmetry that compactifies the scalars, see \cite{Villain}.} gives the action
\begin{equation}
    S_{op-combined} = \frac{iN}{2\pi}\sum \varphi^{xy}\Delta_\tau \Delta_x (\phi_+-\phi_-).
\end{equation}
This gives the same fusion rule as the continuum theory.  

If the symmetry is a defect, the initial boundary term is
\begin{equation}
    S_\partial = iN\sum \varphi^{xy} \Delta_x \hat{n}_\tau.
\end{equation}
After summing by parts, we have
\begin{equation}
    S_{defect} = \frac{iN}{2\pi}\sum \phi (-2\pi \Delta_x n_\tau + \Delta_x \Delta_y \varphi^{xy}) + iN\sum \varphi^{xy} \Delta_x \hat{n}_{\tau}. 
\end{equation}
In the continuum, this becomes the BF type term we asserted earlier.  A similar discussion to the operator case yields the same fusion rule as the continuum theory.

Since the defect has Dirichlet boundary conditions and the $\mathbb{Z}_N$ gauge field is flat, we can deform the defect by a cube without changing any correlation functions.  Thus, we refer to the interface as a symmetry.  The necessity of the square stems from the sensitivity of the $\mathbb{Z}_N$ gauge theory to a preferred foliation structure and accompanying lattice regularization.  This preference is exotic, but perhaps not unexpected of a theory with fracton-related phenomena.
\section{Duality Symmetry in the XY Cube Model}
In this section we construct and analyze the non-invertible duality symmetry in the continuum limit of the XY Cube model.  We also discuss it on the lattice, using a modified Villain model.

\subsection{Review of the Theory}
We begin by reviewing the theory.  Many more details can be found \cite{MoreExotic}.  One continuum presentation of the XY Cube Model is in terms of a scalar $\phi$ subject to the identification $\phi\sim\phi + 2\pi n^x(x) + 2\pi n^y(y) +2\pi n^z(z)$, where $n^i(x^i)$ are integer-valued functions of the appropriate coordinate.  Its Lagrangian is
\begin{equation}
    \mathcal{L} = \frac{\mu_0}{2}(\partial_\tau \phi)^2 + \frac{1}{2\mu}(\partial_x \partial_y \partial_z \phi)^2.
\end{equation}
Thanks to the identification on $\phi$, operators such as $\phi$ and $\partial_i  \phi$ are not well defined, but operators such as $\exp[i\phi]$ and $\partial_x \partial_y \partial_z \phi$ are.  The theory has two symmetries that impact the following.  The first is a $U(1)$ momentum quadrupole symmetry that shifts $\phi$.  It follows from the equation of motion:
\begin{equation}
    \partial_\tau J^\tau = \partial_x \partial_y \partial_z J^{xyz},
\end{equation}
where
\begin{equation}
    J^\tau = i\mu_0 \partial^\tau \phi ; J^{xyz} = i\frac{1}{\mu}\partial^x \partial^y \partial^z \phi .
\end{equation}
The theory also has a $U(1)$ winding quadrupole symmetry with currents
\begin{equation}
    J_\tau^{xyz} = \frac{1}{2\pi}\partial^x\partial^y\partial^z \phi ; J = \frac{1}{2\pi} \partial^\tau \phi,
\end{equation}
which obey the obvious continuity equation:
\begin{equation}
    \partial_\tau J_\tau^{xyz} = \partial^x \partial^y \partial^z J.
\end{equation}
There might be more symmetries if we place the theory on a peculiar enough manifold, but we do not consider such cases in this paper.

A second presentation of the theory is in terms of a compact field $\phi^{xy}$ subject to the identification $\phi^{xyz} \sim \phi^{xyz} + 2\pi(n^x(x) + n^y(y) + n^z(z))$, where $n^i(x^i)$ are as above.  Its Lagrangian is
\begin{equation}
    \mathcal{L} = \frac{\Tilde{\mu}_0}{2}(\partial_\tau \phi^{xyz})^2 + \frac{1}{2\Tilde{\mu}}(\partial_x \partial_y \partial_z \phi^{xyz})^2.
\end{equation}
Identical comments about which operators are well defined apply here.  If one explicitly dualizes the $\phi$ presentation of the theory, one can show that the tilded parameters are related to their untilded counterparts as 
\begin{equation}
    \Tilde{\mu}_0 = \frac{\mu}{4\pi^2}; \Tilde{\mu} = 4\pi^2 \mu_0.
\end{equation}
The theory has two symmetries that we will discuss.  It has a dual $U(1)$ momentum quadrupole symmetry.  It follows from the equation of motion:
\begin{equation}
    \partial_\tau J_\tau^{xyz} = \partial^x \partial^y \partial^z J,
\end{equation}
where
\begin{equation}
    J_\tau^{xyz} = i\Tilde{\mu}_0 \partial_\tau \phi^{xyz} ; J = i\frac{1}{\Tilde{\mu}}\partial_x\partial_y\partial_z \phi^{xyz}.
\end{equation}
The theory also has a dual $U(1)$ winding quadrupole symmetry with currents
\begin{equation}
    J_\tau = \frac{1}{2\pi}\partial_x\partial_y\partial_z \phi^{xyz}; J^{xyz} =\frac{1}{2\pi}\partial_\tau \phi^{xyz} 
\end{equation}
satisfying the obvious continuity equation:
\begin{equation}
    \partial_\tau J_\tau^{xyz} = \partial_x \partial_y \partial_z J.
\end{equation}
Analogizing the currents that furnish the same representation of $S_4$ gives the duality relation:
\begin{equation}
    \partial_x \partial_y \partial_z \phi = i\frac{\mu}{2\pi}\partial_\tau \phi^{xyz}; \partial_\tau \phi = \frac{i}{2\pi \mu_0}\partial_x \partial_y \partial_z \phi^{xyz}.
\end{equation}
Thus, we see that this duality is similar to the well known T-duality of the $c=1$ compact boson.
\subsection{Constructing the Symmetry}
We now move to construct the symmetry from half gauging and examine its interaction with other operators and defects in the theory.  We construct the symmetry by gauging a $\mathbb{Z}_N$ subgroup of the $U(1)$ momentum quadrupole symmetry.  This restricts the range of $\phi$ so that we can write an identical looking Lagrangian by defining $\hat{\phi} = N\phi$:
\begin{equation}
    \mathcal{L} = \frac{\mu_0}{2N^2}(\partial_\tau \phi)^2 + \frac{1}{2\mu N^2}(\partial_x\partial_y\partial_z \phi)^2
\end{equation}
Thus, we see that the effect of gauging is to map $\mu_0$ to $\frac{\mu_0}{N^2}$ and $\mu$ to $\mu N^2$.  Remarkably, this can be undone with T-duality for any self-dual values of $\mu$ and $\mu_0$.  Let us now gauge $\mathbb{Z}_N$ in detail.  The appropriate Lagrangian is:
\begin{equation}
    \mathcal{L} = \frac{\mu_0 }{2}(\partial_\tau \phi - A_\tau)^2 + \frac{2}{\mu}(\partial_x\partial_y \partial_z\phi - A_{xyz})^2 + \frac{iN}{2\pi}\phi^{xyz}(\partial_\tau A_{xyz} - \partial_x \partial_y \partial_z A_\tau).
\end{equation}
Here, $A_\tau$ and $A_{xyz}$ are $U(1)$ gauge fields that couple to $J^\tau$ and $J^{xyz}$, respectively.  Thanks to the continuity equation, they have the redundancy:
\begin{equation}
    A_\tau \sim A_\tau + \partial_\tau \alpha; A_{xyz} \sim A_{xyz} + \partial_x \partial_y \partial_z \alpha.
\end{equation}
Integrating out $\phi^{xyz}$ imposes the constraint:
\begin{equation}
    \frac{N}{2\pi}(\partial_\tau A_{xyz} - \partial_x \partial_y \partial_z A_\tau) = 0
\end{equation}
which implies that the gauge fields are $\mathbb{Z}_N$ gauge fields.  Adding the term is the analog of adding a BF term to gauge a $\mathbb{Z}_N$ subgroup of a $U(1)$ symmetry.  The gauged theory has the symmetry defect:
\begin{equation}
    \eta_\tau(x,y,z) = \exp[i\oint d\tau A_\tau] ; \eta_\tau(x,y,z)^N = 1,
\end{equation}
with the second equality following from the equation of motion.  It also has the symmetry operator
\begin{multline}
    \eta_{xyz}(x_1,x_2,y_1,y_2) = \exp[i\int_{x_1}^{x_2}dx\int_{y_1}^{y_2}dy\oint dz A_{xyz}];\\ \eta_{xyz}(y_1,y_2,z_1,z_2) = \exp[i\int_{y_1}^{x_y}dy\int_{z_1}^{z_2}dz\oint dy A_{xyz}];\\  \eta_{xyz}(z_1,z_2,x_1,x_2) = \exp[i\int_{z_1}^{z_2}dz\int_{x_1}^{x_2}dx\oint dz A_{xyz}];\\ \eta_{xyz}(x_1,x_2,y_1,y_2)^N =\eta_{xyz}(y_1,y_2,z_1,z_2)^N =\eta_{xyz}(z_1,z_2,x_1,x_2)^N = 1,
\end{multline}
with the second equality following from the equation of motion.  We interpret these as dual symmetries to the momentum dipole symmetry.  To construct the noninvertible symmetry, we impose Dirichlet boundary conditions on $(A_\tau,A_{xyz})$ along a codimension 1 surface $\mathcal{S}$.  We call the symmetry constructed in this way $\mathcal{D}(\mathcal{S})$.   As before, we treat the symmetry operator and symmetry defect separately.

\subsection{Symmetry Operator}
For simplicity, we consider an operator along $\tau=0$.  The action of the system with the operator is 
\begin{multline}
    S = \int_{\tau\leq 0}d\tau dx dy dz [\frac{\mu_0 }{2}(\partial_\tau \phi_-)^2 + \frac{1}{2\mu}(\partial_x\partial_y \partial_z\phi_-)^2] - \frac{iN}{2\pi}\int_{\tau=0}dx dy dz[\phi_- \partial_x \partial_y \partial_z \phi_+] \\ + \int_{\tau \geq 0} d\tau dx dy dz [\frac{\mu_0 }{2}(\partial_\tau \phi_+)^2 + \frac{1}{2\mu}(\partial_x\partial_y \partial_z \phi_+)^2]
\end{multline}
Why is this the symmetry operator?  In order to preserve the variational principle that is present without the operator, we need to demand 
\begin{equation}
    -\mu_0\partial_\tau \phi_- = \frac{iN}{2\pi}\partial_x \partial_y \partial_z \phi_+
\end{equation}
at $\tau=0$.  Applying the T-duality formula gives
\begin{equation}
      \partial_\tau \phi^{xyz}_+= \frac{1}{N}\partial_\tau \phi_- = -\frac{2\pi i}{\mu}\partial_x \partial_y \partial_x \phi_+.
\end{equation}
We interpret the first equality as gauging $\mathbb{Z}_N$ and appropriately shrinking the scalar and the second as applying the duality.  Thus, this operator does what we want.  Let us see what happens when we fuse an operator and its orientation reversal.  The action for two operators is
\begin{multline}
    S = \int_{\tau\leq 0}d\tau dx dy dz[\frac{\mu_0 }{2}(\partial_\tau \phi_-)^2 + \frac{1}{2\mu}(\partial_x\partial_y\partial_z\phi_-)^2] - \frac{iN}{2\pi}\int_{\tau =0}dx dy dz[\phi_- \partial_x \partial_y \partial_z \phi_I] \\ + \int_{\epsilon \geq \tau \geq 0} d\tau dx dy dz[\frac{\mu_0 }{2}(\partial_\tau \phi_I)^2 + \frac{1}{2\mu}(\partial_x\partial_y\partial_z\phi_I)^2] - \frac{iN}{2\pi}\int_{\tau = \epsilon} dx dy dz [\phi_I \partial_x \partial_y \partial_z \phi_+] \\+ \int_{\tau\geq \epsilon} d\tau dx dy dz[\frac{\mu_0 }{2}(\partial_\tau \phi_+)^2 + \frac{1}{2\mu}(\partial_x\partial_y\partial_z\phi_+)^2].
\end{multline}
Fusion corresponds to taking $\epsilon \rightarrow 0$, in which case action on the operator becomes
\begin{equation}
    S_{op} = \frac{iN}{2\pi}\int dx dy dz \phi_I \partial_x \partial_y\partial_z(\phi_+ - \phi_-).
\end{equation}  
To interpret this fusion rule, note that integrating out $\phi_I$ imposes
\begin{equation}
    \partial_x \partial_y\partial_z (\phi_+-\phi_-) =0; \phi_+-\phi_- \in \frac{2\pi}{N}\mathbb{Z}.
\end{equation}
  Thus, the product operator is the action of $\eta_{xyz}$ to some power.  We have uncovered the fusion rule
\begin{multline}
    \mathcal{D} \times \Bar{\mathcal{D}} \\= \sum_k\sum_{i=1}^N (\eta_{xyz}(x_k,x_{k+1},y_k,y_{k+1})^i+\eta_{xyz}(y_k,y_{k+1},z_k,z_{k+1})^i+\eta_{xyz}(z_k,z_{k+1},x_k,x_{k+1})^i).
\end{multline}
Note that this gauges the a $\mathbb{Z}_N$ subgroup of the $U(1)$ momentum dipole symmetry on the operator.  Thus, we can view the right hand side as a condensation defect obtained from higher gauging.  It requires a lattice regularization for the sum to make sense.  
\subsection{Symmetry Defect}
For simplicity, we consider a defect along $z=0$.  The action of the system with the operator is 
\begin{multline}
    S = \int_{z\leq 0}d\tau dx dy dz [\frac{\mu_0 }{2}(\partial_\tau \phi_L)^2 + \frac{1}{2\mu}(\partial_x\partial_y\partial_z\phi_L)^2] + \frac{iN}{2\pi}\int_{z=0}d\tau dx dy[\phi_L \partial_\tau \partial_x \partial_y \phi_U] \\ + \int_{z \geq 0} d\tau dx dy dz[\frac{\mu_0 }{2}(\partial_\tau \phi_U)^2 + \frac{1}{2\mu}(\partial_x\partial_y \partial_z\phi_U)^2]
\end{multline}
To see that this is the desired defect, note that, in order to preserve the variational principle, we need to require
\begin{equation}
    \frac{iN}{2\pi} \partial_\tau \partial_x \partial_y \phi_U = \frac{1}{2\mu} \partial_x^2 \partial_y^2 \partial_z\phi_L
\end{equation}
at $z=0$.  Applying the T-duality formula gives
\begin{equation}
    \partial_x^2 \partial_y^2 \partial_z \phi_U^{xyz} = \frac{1}{N}\partial_x^2 \partial_y^2 \partial_z \phi_L =-2\pi i \mu_0\partial_\tau \partial_x \partial_y \phi_U .
\end{equation}
We interpret the first equality as gauging $\mathbb{Z}_N$ and appropriately shrinking the scalar and the second as applying the duality.  Therefore, this defect does what we want.  Let us see what happens when we fuse a defect and its orientation reversal.  The action for two defects is 
\begin{multline}
    S = \int_{z \leq 0}d\tau dx dy dz[\frac{\mu_0}{2}(\partial_\tau \phi_L)^2 + \frac{1}{2\mu}(\partial_x\partial_y\partial_z\phi_L)^2] - \frac{iN}{2\pi}\int_{z=0} d\tau dx dy [\phi_L \partial_\tau \partial_x\partial_y \phi_I]\\ + \int_{\epsilon \geq z \geq 0}d\tau dx dy dz[\frac{\mu_0}{2}(\partial_\tau \phi_I)^2 + \frac{1}{2\mu}(\partial_x\partial_y\partial_z\phi_I)^2] - \frac{iN}{2\pi}\int_{z=\epsilon} d\tau dx dy [\phi_I \partial_\tau \partial_x\partial_y \phi_U] \\ + \int_{z \geq \epsilon}d\tau dx dy dz [\frac{\mu_0}{2}(\partial_\tau \phi_U)^2 + \frac{1}{2\mu}(\partial_x\partial_y\partial_z\phi_U)^2]. 
\end{multline}
Fusion corresponds to taking $\epsilon \rightarrow 0$.  The resulting defect action is
\begin{equation}
    S = \frac{iN}{2\pi}\int_{z=0} d\tau dx dy [\phi_I\partial_\tau\partial_x\partial_y(\phi_U - \phi_L)].
\end{equation}
To understand this, note that integrating out $\phi_I$ imposes
\begin{equation}
    \partial_\tau \partial_x\partial_y(\phi_U-\phi_L) =0; \phi_U-\phi_L \in \frac{2\pi}{N}\mathbb{Z}.
\end{equation}
  Thus, the product defect is an action of $\eta_\tau$ to some power.  We have uncovered the fusion rule:
\begin{equation}
    \mathcal{D} \times \Bar{\mathcal{D}} = \sum_{(x,y,z)} \sum_{i=1}^N \eta_\tau (x,y,z)^N 
\end{equation}
Note that we require a lattice regularization to make the above sum discrete.  The right hand side gauges a $\mathbb{Z}_N$ subgroup of the $U(1)$ momentum dipole symmetry along the defect.  Thus, we can view it as a condensation defect from higher gauging.  
\subsection{Modified Villain Construction}

In this subsection, we construct both the non-invertible operator and non-invertible defect on a Euclidean lattice.  This construction deduces the form of the operator and defect actions rather than asserting them.  Again, we follow the strategy in \cite{NonInvDual}.  The model lives on the Euclidean hypercubic lattice $\mathbb{Z}_{L_\tau} \times \mathbb{Z}_{L_x} \times \mathbb{Z}_{L_y}\times \mathbb{Z}_{L_z}$.  The action takes the form:
\begin{multline}
    S = \frac{\mu_0}{2}\sum_{\tau-links} (\Delta_\tau \phi - 2\pi n_\tau)^2 + \frac{1}{2\mu}\sum_{xyz-cubes}(\Delta_x \Delta_y \Delta_z \phi - 2\pi n_{xyz})^2 \\+ i\sum_{dual-sites}\phi^{xyz}(\Delta_\tau n_{xyz} - \Delta_x\Delta_y\Delta_z n_\tau)
\end{multline}
Here, $\Delta_i$ is the lattice derivative along a link in the $i^{th}$ direction, $\phi$ is a real-valued variable on a site, $n_\tau$ is an integer-valued variable on a $\tau$ link, $n_{xyz}$ is an integer-valued variable on an xyz cube, and $\phi^{xyz}$ is a real-valued variable on a dual site.  The fields are subject to the following identifications:
\begin{equation}
    \phi \sim \phi + 2\pi k
\end{equation}
\begin{equation}
    n_\tau \sim n_\tau + \Delta_\tau k
\end{equation}
\begin{equation}
    n_{xyz} \sim n_{xyz} + \Delta_x \Delta_y \Delta_z k,
\end{equation}
\begin{equation}
    \phi^{xyz} \sim \phi^{xyz} + 2\pi \Tilde{k}^{xyz}
\end{equation}
where $k$ and $\Tilde{k}^{xyz}$ are integer-valued variables on (dual) sites.  The first and fourth identifications compactify the target space.  If we view $n$ as a gauge field, then $\phi^{xyz}$ enforces a flatness constraint.  Applying the Poisson resummation formula for both $n_\tau$ and $n_{xyz}$ above gives a dual presentation of the model, with action:
\begin{multline}
    S = \frac{\Tilde{\mu}_0}{2}\sum_{dual-\tau-links} (\Delta_\tau \phi^{xyz} - 2\pi \Tilde{n}^{xyz}_\tau)^2 + \frac{2}{\Tilde{\mu}}\sum_{dual-xyz-cubes}(\Delta_x \Delta_y \Delta_z \phi^{xyz} - 2\pi \Tilde{n})^2 \\+ i\sum_{sites}\phi(\Delta_\tau \Tilde{n} - \Delta_x \Delta_y \Delta_z \Tilde{n}^{xyz}_\tau ), 
\end{multline}
where
\begin{equation}
    \Tilde{n}^{xyz}_\tau \sim  \Tilde{n}^{xyz}_\tau + \Delta_\tau \Tilde{k}^{xyz}
\end{equation}
\begin{equation}
    \Tilde{n} \sim \Tilde{n} + \Delta_x \Delta_y \Delta_z \Tilde{k}^{xyz}.
\end{equation}
This is a self duality, so everything in the above action has the same interpretation as the $\phi$ action, just on the dual lattice.  

We proceed as in the continuum, beginning by gauging the $\mathbb{Z}_N$ subgroup of the $U(1)$ momentum dipole symmetry.  The action is
\begin{multline}
    S = \frac{\mu_0}{2}\sum_{\tau-links} (\Delta_\tau \phi - 2\pi n_\tau - \frac{2\pi}{N}\Hat{n}_\tau)^2 + \frac{1}{2\mu}\sum_{xyz-cubes}(\Delta_x \Delta_y \Delta_z \phi - 2\pi n_{xyz} - \frac{2\pi}{N}\hat{n}_{xyz}) \\+ i\sum_{dual-sites}\phi^{xyz}(\Delta_\tau n_{xyz} - \Delta_x\Delta_y\Delta_z n_\tau + \frac{1}{N}\Delta_\tau \hat{n}_{xyz} -\frac{1}{N}\Delta_x \Delta_y\Delta_z \hat{n}_\tau) \\+ \frac{2\pi i}{N}\sum_{dual-sites} \hat{m}^{xyz}(\Delta_\tau \hat{n}_{xyz} -\Delta_x \Delta_y \Delta_z \hat{n}_\tau).
\end{multline}
The fields are subject to the identifications:
\begin{equation}
    \phi \sim \phi + 2\pi k + \frac{2\pi q}{N}
\end{equation}
\begin{equation}
    n_\tau \sim n_\tau + \Delta_\tau k - l_\tau
\end{equation}
\begin{equation}
    n_{xyz} \sim n_{xyz} + \Delta_x \Delta_y \Delta_z k - l_{xyz}
\end{equation}
\begin{equation}
    \phi^{xyz} \sim \phi^{xyz} + 2\pi \Tilde{k}^{xyz}
\end{equation}
\begin{equation}
    \Hat{n}_\tau \sim \Hat{n}_\tau + \Delta_\tau q + Nl_\tau
\end{equation}
\begin{equation}
    \Hat{n}_{xyz} \sim \Hat{n}_{xyz} + \Delta_x \Delta_y \Delta_z q + N l_{xyz}
\end{equation}
\begin{equation}
    \Hat{m}^{xyz} \sim \Hat{m}^{xyz} - \Tilde{k}^{xyz} + N q^{xyz}.
\end{equation}
Above, $q$ is an integer-valued field on a site, $q^{xyz}$ is an integer valued field on a dual site, $l_\tau$ is an integer-valued field on a $\tau$ link, $l_{xyz}$ is an integer-valued field on an $xyz$ cube, $\hat{n}_\tau$ is an integer-valued field on a $\tau$ link, $\hat{n}_{xyz}$ is an integer-valued field on an $xyz$ cube, $\hat{m}^{xyz}$ is an integer-valued field on a dual site.  We can gauge away $\Hat{m}^{xyz}$, $n_\tau$, and $n_{xyz}$ and redefine variables: 
\begin{equation}
    \varphi := N \phi ; \varphi^{xyz} := \frac{\phi^{xyz}}{N}.
\end{equation}
This yields the action:
\begin{multline}
    S = \frac{\mu_0}{2N^2} \sum_{\tau-links} (\Delta_\tau \varphi - 2\pi \hat{n}_\tau)^2 + \frac{1}{2N^2 \mu}\sum_{xyz-cubes}(\Delta_x \Delta_y \Delta_z \varphi - 2\pi \hat{n}_{xyz})^2 \\ +i \sum_{dual-sites} \varphi^{xyz}(\Delta_\tau \hat{n}_{xyz} - \Delta_x \Delta_y \Delta_z \hat{n}_\tau).
\end{multline}
Note that gauging $\mathbb{Z}_N$ indeed rescales the fields as anticipated from our continuum analysis.

To construct the symmetry, we simply apply the above to the theory on half of spacetime.  Note that we acquire a boundary term in killing the $n$s.  This term is different for the operator and the defect, so we will discuss it on a case by case basis.  Let's start with what we can say independent of whether the symmetry is an operator or defect.  In particular, we can apply the Poisson resummation formula to the theory on the half of spacetime on which we gauge it.  The resulting action is
\begin{multline}
    S = \frac{\Tilde{\mu}_0}{2}\sum_{\tau-links} (\Delta_\tau \varphi^{xyz} - 2\pi \Tilde{n}_\tau)^2 + \frac{1}{2\Tilde{\mu}}\sum_{xyz-cubes}(\Delta_x \Delta_y \Delta_z \varphi^{xyz} - 2\pi \Tilde{n}_{xyz})^2 \\ + \frac{i}{2\pi}\sum \Delta_\tau \varphi (2\pi \Tilde{n}_{xyz} - \Delta_x \Delta_y \Delta_z \varphi^{xyz}) + \frac{i}{2\pi}\sum \Delta_x \Delta_y \Delta_z \varphi (2\pi \Tilde{n}_\tau - \Delta_\tau \varphi^{xyz}) + S_\partial
\end{multline}
We can sum by parts to obtain the $\varphi^{xyz}$ presentation of the theory.  Since this clearly depends on whether we have an operator or a defect, we will treat them in turn.  If the symmetry is an operator, the initial boundary term is
\begin{equation}
    S_\partial = -iN \sum \varphi^{xyz}n_{xyz}.
\end{equation}
After summing by parts, we have
\begin{equation}
    S_{operator} = -\frac{iN}{2\pi}\sum \phi (2\pi \Tilde{n}_{xyz}-\Delta_x\Delta_y\Delta_z \varphi^{xyz}) - iN\sum \varphi^{xyz} n_{xyz}.
\end{equation}
In the continuum, this becomes the BF type term we asserted earlier.  Just as in the previous section, one can show that this obeys the same fusion rule as its continuum counterpart.  If the symmetry is a defect, the initial boundary term is
\begin{equation}
    S_\partial = iN\sum \varphi^{xyz} \Delta_x \Delta_y n_\tau.
\end{equation}
After summing by parts, we have
\begin{equation}
    S_{defect} = \frac{iN}{2\pi}\sum \phi (-2\pi \Delta_x \Delta_y n_\tau + \Delta_x \Delta_y \Delta_\tau \varphi^{xyz}) + iN\sum \varphi^{xyz} \Delta_x \Delta_y n_\tau. 
\end{equation}
In the continuum, this becomes the BF type term we asserted earlier.  Just as in the previous section, one can show that this obeys the same fusion rule as its continuum counterpart.

Since the $(A_\tau, A_{xyz})$ is flat and subject to Dirichlet boundary conditions, we can deform the defect, so that it is a symmetry.  Note that this theory is sensitive to the lattice cutoof, so we can only deform by a hypercube.
\section{Duality Symmetry in the $1+1 D$ Compact Lifshitz Theory}
In this section, we construct and analyze a non-invertible duality symmetry in a modified Villain lattice model of the $1+1 D$ compact Lifshitz theory.  As detailed in \cite{TimelikeSymmetry}, the relationship between the symmetries in the Villain model and the continuum is subtle, and depends on the choice of continuum limit.  Our discussion in the Villain model does not straightforwardly generalize to any of known limits - we leave its fate in the continuum to future work. 
\subsection{Review of the Theory}
In this subsection, we review the Villain lattice model of the $1+1D$ compact Lifshitz theory, following \cite{TimelikeSymmetry}. We will work on a square lattice $\mathbb{Z}_{L_\tau}\times \mathbb{Z}_{L_x}$.  The modified Villain action is
\begin{multline}
    S = \frac{\mu_0}{2}\sum_{\tau-links}(\Delta_\tau \phi - 2\pi n_\tau)^2 + \frac{1}{2\mu}\sum_{sites}(\Delta_x^2 \phi - 2\pi n_{xx})^2\\ + i\sum_{\tau-links} \Tilde{\phi}(\Delta_\tau n_{xx} -\Delta_x^2 n_\tau). 
\end{multline}
Here, $\phi$ is a real-valued variable on a site, $n_\tau$ is an integer-valued variable on a $\tau$ link, $n_{xx}$ is an integer-valued variable on a site, and $\Tilde{\phi}$ is a real-valued variable on a $\tau$ link.  The fields are subject to the following gauge identifications:
\begin{equation}
    \phi \sim \phi + 2\pi k
\end{equation}
\begin{equation}
    n_\tau \sim n_\tau + \Delta_\tau k
\end{equation}
\begin{equation}
    n_{xx} \sim n_{xx} + \Delta_x^2 k
\end{equation}
\begin{equation}
    \Tilde{\phi} \sim \Tilde{\phi}+2\pi \Tilde{k}.
\end{equation}
The first and fourth identifications compactify the target space.  The second two motivate viewing $n_\tau$ and $n_{xx}$ as components of a gauge field.  $\Tilde{\phi}$ enforces a flatness constraint on said gauge field.  Let us discuss the global symmetries of the theory.  First, there is a $U(1)$ momentum symmetry that maps $\phi \rightarrow \phi + c$, where $c \in \mathbb{R}$.  The identification on $\phi$ makes this $U(1)$.  This symmetry follows from the equation of motion for $\phi$:
\begin{equation}
    \Delta_\tau J_\tau = \Delta_x^2 J_{xx},
\end{equation}
 where 
\begin{equation}
     J_\tau = i\mu_0 (\Delta_\tau \phi - 2\pi n_\tau); J_{xx} = \frac{i}{\mu}(\Delta_x^2 \phi - 2\pi n_{xx}).
\end{equation}
Second, there is a $U(1)$ winding symmetry that maps $\Tilde{\phi} \rightarrow \Tilde{\phi} + \Tilde{c}$.  The identification on $\Tilde{\phi}$ makes this symmetry $U(1)$.  This symmetry follows from the equation of motion for $\Tilde{\phi}$:
\begin{equation}
    \Delta_\tau \Tilde{J}_\tau = \Delta_x^2 \Tilde{J}_{xx},
\end{equation}
where
\begin{equation}
    J_\tau = \frac{1}{2\pi}(\Delta_x^2 \phi - 2\pi n_{xx}); J_{xx} = \frac{1}{2\pi}(\Delta_\tau \phi - 2\pi n_\tau).
\end{equation}
Third, there is a $\mathbb{Z}_{L_x}$ momentum dipole symmetry that maps $\phi \rightarrow \phi + 2\pi m \frac{x}{L_x}$ and $n_{xx} \rightarrow n_{xx} + m(\delta_{x0} - \delta_{x L_x-1})$, where $m \in 0,...,L_x-1$ and $x \in [0,L_x)$.  This is a $\mathbb{Z}_{L_x}$ symmetry because $m \in L_z \mathbb{Z}$ is part of the identification.  
Fourth, there is a $\mathbb{Z}_{L_x}$ winding dipole symmetry that maps $\Tilde{\phi} \rightarrow \Tilde{\phi} + 2\pi \Tilde{m}\frac{x}{L_x}$, where $\Tilde{m} \in 0,...,L_x-1$ and $x \in [0,L_x)$.  This is a $\mathbb{Z}_{L_x}$ symmetry because $\Tilde{m}\in L_z\mathbb{Z}$ is part of the identification.  

By applying the Poisson resummation formula, one can obtain a dual presentation of the theory.  The dual presentation is
\begin{multline}
     S = \frac{\Tilde{\mu}_0}{2}\sum_{sites}(\Delta_\tau \Tilde{\phi} - 2\pi \Tilde{n}_\tau)^2 + \frac{1}{2\mu}\sum_{\tau-links}(\Delta_x^2 \Tilde{\phi} - 2\pi \Tilde{n}_{xx})^2\\ + i\sum_{sites} \phi(\Delta_\tau \Tilde{n}_{xx} -\Delta_x^2 \Tilde{n}_\tau).
\end{multline}
This duality obviously exchanges momentum and winding symmetries.  The old and new parameters are related as always.  Analogizing the symmetries that transform the same way gives the duality relation:
\begin{equation}
    i(\Delta_x^2\phi - 2\pi n_{xx}) = \frac{\mu}{2\pi}(\Delta_\tau \Tilde{\phi} - 2\pi \Tilde{n}_\tau) ; i(\Delta_\tau \phi - 2\pi n_\tau) = \frac{1}{2\pi \mu_0}(\Delta_x^2\Tilde{\phi} - 2\pi \Tilde{n}_{xx}).
\end{equation}
\subsection{Constructing the Symmetry}
We construct the symmetry as usual - by gauging a $\mathbb{Z}_N$ subgroup of the $U(1)$ momentum symmetry on half of the spacetime.  After appropriately adding gauge fields, the action is
\begin{multline}
    S = \frac{\mu_0}{2}\sum_{\tau-links}(\Delta_\tau \phi - 2\pi n_\tau - \frac{2\pi}{N}\hat{n}_\tau)^2 + \frac{1}{2\mu}\sum_{sites}(\Delta_x^2 \phi - 2\pi n_{xx} - \frac{2\pi}{N}\hat{n}_{xx})^2\\ + i\sum_{\tau-links} \Tilde{\phi}(\Delta_\tau n_{xx} -\Delta_x^2 n_\tau + \frac{1}{N}\Delta_\tau \hat{n}_{xx} - \frac{1}{N}\Delta_x^2 \hat{n}_\tau) \\ + \frac{2\pi i}{N} \sum_{\tau-links} \Tilde{m}(\Delta_\tau \hat{n}_{xx} - \Delta_x^2 \hat{n}_\tau).
\end{multline}
The fields are subject to the identifications:
\begin{equation}
    \phi \sim \phi + 2\pi k + \frac{2\pi q}{N}
\end{equation}
\begin{equation}
    n_\tau \sim n_\tau + \Delta_\tau k - l_\tau
\end{equation}
\begin{equation}
    n_{xx} \sim n_{xx} + \Delta_x^2 k - l_{xx}
\end{equation}
\begin{equation}
    \Tilde{\phi} \sim \Tilde{\phi}+2\pi \Tilde{k}.
\end{equation}
\begin{equation}
    \hat{n}_\tau \sim \hat{n}_\tau + \Delta_\tau q + Nl_\tau
\end{equation}
\begin{equation}
    \hat{n}_{xx} \sim \hat{n}_{xx} + \Delta_x^2 q + Nl_{xx}
\end{equation}
\begin{equation}
    \Tilde{m} \sim \Tilde{m} - \Tilde{k} + N \Tilde{q}.
\end{equation}
Above, $q$ is an integer-valued variable on a site, $l_\tau$ is an integer-valued variable on a $\tau$ link, $l_{xx}$ is an integer-valued variable on a site, $\Tilde{m}$ is an integer-valued variable on a $\tau$ link, and $\Tilde{q}$ is an integer-valued variable on a $\tau$ link.   We can gauge away $n_\tau$, $n_{xx}$, and $\Tilde{m}$ and redefine fields:
\begin{equation}
    \varphi := N \phi ; \Tilde{\varphi} := \frac{\Tilde{\phi}}{N}, 
\end{equation}
leaving
\begin{multline}
    S = \frac{\mu_0}{2N^2}\sum_{\tau-links}(\Delta_\tau \varphi  - 2\pi\hat{n}_\tau)^2 + \frac{1}{2\mu N^2}\sum_{sites}(\Delta_x^2 \varphi  - 2\pi\hat{n}_{xx})^2\\ + iN\sum_{\tau-links} \Tilde{\varphi}(\Delta_\tau \hat{n}_{xx} - \Delta_x^2 \hat{n}_\tau),
\end{multline}
which is the original theory with parameters $\frac{\mu_0}{N^2}$ and $\mu N^2$.  Thus, for the same special values of parameters, we can undo the combination of self-duality gives the same action.  

To use this to make the symmetry, we apply the above gauging procedure on half of the spacetime.  Along the interface, we set $\hat{n}_\tau = \hat{n}_{xx} = 0$.  There is a boundary term from killing $(n_\tau,n_{xx})$.  It differs based on whether the symmetry is an operator or a defect, so we will address the specifics on a case by case basis.  What we will do initially is apply the duality to the gauged theory to return to the original theory.  To do so, apply the Poisson resummation formula to $(\hat{n}_\tau,\hat{n}_{xx})$, yielding
\begin{multline}
    S = \frac{\Tilde{\mu_0}}{2}\sum_{\tau-links}(\Delta_\tau \Tilde{\varphi} - 2\pi \Tilde{n}_\tau)^2 + \frac{2}{\Tilde{\mu}}\sum_{sites} (\Delta_x^2\Tilde{\varphi} - 2\pi \Tilde{n})^2 \\ + \frac{i}{2\pi}\sum_{\tau-links} \Delta_\tau \varphi (2\pi \Tilde{n} - \Delta_x^2\Tilde{\varphi}) + \frac{i}{2\pi}\sum_{\tau-links} \Delta_x^2\varphi(2\pi\Tilde{n}_\tau - \Delta_\tau \Tilde{\varphi}) + S_\partial
\end{multline}
\subsection{Symmetry Operator}
We consider a symmetry operator at $\tau =0$.  In this case, the boundary term is
\begin{equation}
    S_\partial = -iN\sum_{\tau-links;\hat{\tau}=0} \Tilde{\varphi} n_{xx}.
\end{equation}
So that, after summing by parts, we have the operator action
\begin{equation}
    S_{operator} = -\frac{iN}{2\pi}\sum_{\tau-links;\hat{\tau}=0} \phi (2\pi \Tilde{n} -\Delta_x^2 \Tilde{\varphi})  -iN\sum_{\tau-links;\hat{\tau}=0} \Tilde{\varphi} n_{xx}.
\end{equation}
Let us now consider the fusion of an operator and its opposite.  Following the same procedure as above for the dual theory on $\tau \leq 0$ gives
\begin{equation}
    \Bar{S}_{operator} = -\frac{iN}{2\pi}\sum_{\tau-links;\hat{\tau}=0} \Tilde{\varphi}(\Delta_x^2 \phi - 2\pi n_{xx}) + iN\sum_{\tau-links;\hat{\tau}=0} \phi \Tilde{n}
\end{equation}
so that, after gauging away $n_{xx}$ and $\Tilde{n}$ and fusing the two, we obtain
\begin{equation}
    S_{combined} = \frac{iN}{2\pi}\sum_{\tau-links;\hat{\tau}=0} \Tilde{\varphi}\Delta_x^2(\phi_+-\phi_-) .
\end{equation}
We interpret this as a condensation operator as before.
\subsection{Symmetry Defect}
We consider a symmetry defect at $x=0$.  In this case, the boundary term is 
\begin{equation}
    S_\partial = iN \sum_{x\tau-squares; \hat{x}=0} \Tilde{\varphi} \Delta_x n_\tau.  
\end{equation}
After summing by parts, we obtain the defect action:
\begin{equation}
    S_{defect} = -2\frac{iN}{2\pi}\sum_{x-links; \hat{x}=0} \phi(-2\pi \Delta_x\Tilde{n}_\tau + \Delta_\tau \Delta_x \Tilde{\varphi})+iN \sum_{x\tau-squares; \hat{x}=0} \Tilde{\varphi} \Delta_x n_\tau.
\end{equation}
Let us now consider the fusion of a defect and its opposite.  Following the same procedure as above for the dual theory on $x \leq 0$, we obtain
\begin{equation}
    \Bar{S}_{defect} = 2\frac{iN}{2\pi}\sum_{x\tau-squares; \hat{x}=0} \Tilde{\varphi}(-2\pi\Delta_x n_\tau + \Delta_\tau \Delta_x \phi) - iN\sum_{x-links; \hat{x}=0} \phi \Delta_x\Tilde{n}_\tau.  
\end{equation}
Upon fusing these and gauging away $n_\tau$ and $\Tilde{n}_\tau$, we obtain
\begin{equation}
    S_{combined} = -2\frac{iN}{2\pi}\sum \Tilde{\varphi} \Delta_\tau \Delta_x(\phi_R-\phi_L).
\end{equation}
Just as above, this is a condensation defect.  The difference is that here, one gauges a $\mathbb{Z}_{2N}$ subgroup of the $U(1)$ momentum symmetry on the defect.  We note that the defect is on an adjacent pair of level-x lines.

As demonstrated in the appendix, one can deform the defect, so this duality interface is a symmetry.
\section{Duality Symmetry in the $2+1$ Dimensional Laplacian Lifshitz Theory}
In this section, we construct and analyze the non-invertible duality symmetry in the modified Villain lattice model of the $2+1$ dimensional Laplacian Lifshitz theory.  As in the $1+1$ dimensional theory, the relationship between the Villain theory and the continuum is subtle, as the Villain theory is one of two very different regularizations of the same continuum theory \cite{CptLifshitz}.  We leave a discussion of the fate of this symmetry in the continuum to future work. 
\subsection{Review of the Theory}
In this subsection, we review the Villain lattice model of the $2+1D$ Laplacian Lifshitz theory, following \cite{CptLifshitz}.  We will work on a cubic lattice $\mathbb{Z}_{L_\tau}\times\mathbb{Z}_{L_x}\times\mathbb{Z}_{L_y}$.  The modified Villain action is
\begin{multline}
    S =\frac{\mu_0}{2}\sum_{\tau-links}(\Delta_\tau \phi - 2\pi n_\tau)^2 + \frac{1}{2\mu}\sum_{sites}[(\Delta_x^2 + \Delta_y^2)\phi - 2\pi n]^2 \\ + i\sum_{\tau-links}\Tilde{\phi}[\Delta_\tau n - (\Delta_x^2+\Delta_y^2)n_\tau].
\end{multline}
Above, $\phi$ is a real-valued variable on a site, $n_\tau$ is an integer-valued variable on a $\tau$ link, $n$ is an integer-valued variable on a site, and $\Tilde{\phi}$ is a real-valued variable on a $\tau$ link.  The fields are subject to the following identifications:
\begin{equation}
    \phi \sim \phi + 2\pi k
\end{equation}
\begin{equation}
    n_\tau \sim n_\tau + \Delta_\tau k
\end{equation}
\begin{equation}
    n \sim n + (\Delta_x^2 + \Delta_y^2)k
\end{equation}
\begin{equation}
    \Tilde{\phi} \sim \Tilde{\phi} + 2\pi \Tilde{k}.
\end{equation}
The first and fourth identifications compactify the target space.  The second and third motivate viewing the $n$s as gauge fields.  From this point of view, $\Tilde{\phi}$ is a Lagrange multiplier that forces them to be flat.  Let us discuss the global symmetries of the theory:
First, there is a $U(1)$ momentum symmetry that maps $\phi \rightarrow \phi + c$.  The identification on $\phi$ makes this $U(1)$.  This symmetry follows from the equation of motion for $\phi$:
\begin{equation}
        \Delta_\tau J_\tau = (\Delta_x^2+\Delta_y^2)J,
    \end{equation}
where 
\begin{equation}
    J_\tau = i\mu_0 (\Delta_\tau \phi - 2\pi n_\tau); J = \frac{i}{\mu}[(\Delta_x^2 + \Delta_y^2)\phi - 2\pi n].
\end{equation}
Second, there is a $U(1)$ winding symmetry that maps $\Tilde{\phi} \rightarrow \Tilde{\phi} + \Tilde{c}$.  The identification on $\Tilde{\phi}$ makes this a $U(1)$ symmetry.  This symmetry follows from the equation of motion for $\Tilde{\phi}$:
\begin{equation}
    \Delta_\tau J_\tau = [\Delta_x^2+\Delta_y^2]J,
\end{equation}
where
\begin{equation}
    J_\tau = \frac{1}{2\pi}[(\Delta_x^2 + \Delta_y^2)\phi - 2\pi n] ; J = \frac{1}{2\pi}(\partial_\tau \phi - 2\pi n_\tau).
\end{equation}
Third, there is a $Jac(C_{L_x}\times C_{L_y})$ \footnote{$C_i$ is the cyclic graph with $i$ vertices.  $Jac(C_{L_x}\times C_{L_y})$ is the Jacobian group of $C_{L_x}\times C_{L_y}$.  For details, see \cite{Graph}.} momentum symmetry.  It maps:
\begin{equation}
    \phi \rightarrow \phi + f(x,y)
\end{equation}
\begin{equation}
    n \rightarrow n + \frac{1}{2\pi}(\Delta_x^2 + \Delta_y^2)f(x,y),
\end{equation}
where
\begin{equation}
    (\Delta_x^2 + \Delta_y^2)f(x,y) \in 2\pi \mathbb{Z}
\end{equation}
i.e. $f$ is a circle valued harmonic function on a torus.
Fourth, there is a $Jac(C_{L_x}\times C_{L_y})$ winding symmetry that maps
\begin{equation}
    \Tilde{\phi} \rightarrow \Tilde{\phi} + \Tilde{f}(x,y),
\end{equation}
where $\Tilde{f}(x,y)$ is a circle valued harmonic function on the torus.
By applying the Poisson resummation formula, we can dualize this theory.  The dual presentation is:
\begin{multline}
    S = \frac{\Tilde{\mu}_0}{2}\sum_{\tau-links}(\Delta_\tau \Tilde{\phi} - 2\pi \Tilde{n}_\tau)^2 + \frac{1}{2\Tilde{\mu}}\sum_{sites}[(\Delta_x^2 + \Delta_y^2)\Tilde{\phi} - 2\pi \Tilde{n}]^2 \\ + i\sum_{\tau-links}\phi[\Delta_\tau \Tilde{n} - (\Delta_x^2+\Delta_y^2)\Tilde{n}_\tau]
\end{multline}
The duality obviously exchanges momentum and winding symmetries.  The old and new parameters are related as always.  The duality relation is:
\begin{equation}
    i([\Delta_x^2 + \Delta_y^2]\phi - 2\pi n) = \frac{\mu}{2\pi}(\Delta_\tau \Tilde{\phi} - 2\pi \Tilde{n}_\tau) ; i(\Delta_\tau \phi - 2\pi n_\tau) = \frac{1}{2\pi \mu_0}([\Delta_x^2 + \Delta_y^2]\Tilde{\phi} - 2\pi n)
\end{equation}
\subsection{Constructing the Symmetry}
We construct the symmetry as usual - by gauging a $\mathbb{Z}_N$ subgroup of the $U(1)$ momentum symmetry on half of the spacetime.  After appropriately adding gauge fields, the action is
\begin{multline}
    S = \frac{\mu_0}{2}\sum_{\tau-links}(\Delta_\tau \phi - 2\pi n_\tau - \frac{2\pi}{N}\hat{n}_\tau)^2 + \frac{1}{2\mu}\sum_{sites}[(\Delta_x^2 + \Delta_y^2)\phi - 2\pi n - \frac{2\pi}{N}\hat{n}]^2\\ + i\sum_{\tau links} \Tilde{\phi}[\Delta_\tau n -(\Delta_x^2 + \Delta_y^2)n_\tau + \frac{1}{N}\Delta_\tau \hat{n} - \frac{1}{N}(\Delta_x^2 + \Delta_y^2)\hat{n}_\tau] \\ + \frac{2\pi i}{N} \sum_{\tau-links} \Tilde{m}[\Delta_\tau \hat{n} - (\Delta_x^2 + \Delta_y^2)\hat{n}_\tau].
\end{multline}
The fields are subject to the identifications:
\begin{equation}
    \phi \sim \phi + 2\pi k + \frac{2\pi q}{N}
\end{equation}
\begin{equation}
    n_\tau \sim n_\tau + \Delta_\tau k - l_\tau
\end{equation}
\begin{equation}
    n \sim n + (\Delta_x^2 + \Delta_y^2)k - l
\end{equation}
\begin{equation}
    \Tilde{\phi} \sim \Tilde{\phi}+2\pi \Tilde{k}.
\end{equation}
\begin{equation}
    \hat{n}_\tau \sim \hat{n}_\tau + \Delta_\tau q + Nl_\tau
\end{equation}
\begin{equation}
    \hat{n} \sim \hat{n} + (\Delta_x^2 + \Delta_y^2)q + Nl
\end{equation}
\begin{equation}
    \Tilde{m} \sim \Tilde{m} - \Tilde{k} + N \Tilde{q}.
\end{equation}
Above, $q$ is an integer-valued variable on a site, $l_\tau$ is an integer-valued variable on a $\tau$ link, $l$ is an integer-valued variable on a site, $\hat{n}_\tau$ is an integer-valued variable on a $\tau$ link, $\hat{n}$ is an integer-valued variable on a site, $\Tilde{m}$ is an integer-valued variable on a $\tau$ link, and $\Tilde{q}$ is an integer-valued variable on a $\tau$ link.  We can gauge away $n_\tau$, $n$, and $\Tilde{m}$ and redefine fields:
\begin{equation}
    \varphi := N \phi ; \Tilde{\varphi} := \frac{\Tilde{\phi}}{N}, 
\end{equation}
leaving
\begin{multline}
    S = \frac{\mu_0}{2N^2}\sum_{\tau-links}(\Delta_\tau \varphi  - 2\pi\hat{n}_\tau)^2 + \frac{1}{2\mu N^2}\sum_{sites}[(\Delta_x^2 + \Delta_y^2)\varphi  - 2\pi\hat{n}]^2\\ + i\sum_{\tau-link} \Tilde{\varphi}[\Delta_\tau \hat{n} - (\Delta_x^2+\Delta_y^2) \hat{n}_\tau],
\end{multline}
which is the original theory with parameters $\frac{\mu_0}{N^2}$ and $\mu N^2$.  Thus, for the same special values of parameters, we can undo the combination of self-duality gives the same action.  

To use this to make the symmetry, we apply the above gauging procedure on half of the spacetime.  Along the interface, we set $\hat{n}_\tau = \hat{n} = 0$.  There is a boundary term from killing $(n_\tau,n)$.  It differs based on whether the symmetry is an operator or a defect, so we will address the specifics on a case by case basis.  What we will initially do is apply the duality to the gauged theory to return to the original theory.  To do so, apply the Poisson resummation formula to $(\hat{n}_\tau,\hat{n})$, yielding
\begin{multline}
    S = \frac{\Tilde{\mu_0}}{2}\sum_{\tau-links}(\Delta_\tau \Tilde{\varphi} - 2\pi \Tilde{n}_\tau)^2 + \frac{1}{2\Tilde{\mu}}\sum_{sites} [(\Delta_x^2 + \Delta_y^2)\Tilde{\varphi} - 2\pi \Tilde{n}]^2 \\ + \frac{i}{2\pi}\sum_{\tau-links} \Delta_\tau \varphi [2\pi \Tilde{n} - (\Delta_x^2 + \Delta_y^2)\Tilde{\varphi}] + \frac{i}{2\pi}\sum_{sites} (\Delta_x^2 + \Delta_y^2)\varphi(2\pi\Tilde{n}_\tau - \Delta_\tau \Tilde{\varphi}) + S_\partial
\end{multline}
\subsection{Symmetry Operator}
We consider a symmetry operator at $\tau =0$.  In this case, the boundary term is
\begin{equation}
    S_\partial = -iN\sum_{\tau-links;\hat{\tau}=0} \Tilde{\varphi} n.
\end{equation}
So that, after summing by parts, we have the operator action
\begin{equation}
    S_{operator} = -\frac{iN}{2\pi}\sum_{\tau-links;\hat{\tau}=0} \phi [2\pi \Tilde{n} -(\Delta_x^2 +\Delta_y^2)\Tilde{\varphi}]  -iN\sum_{\tau-links;\hat{\tau}=0} \Tilde{\varphi} n.
\end{equation}
Let us now consider fusing this operator and its opposite.  Following the same procedure as above of the dual theory on $\tau \leq 0$ gives 
\begin{equation}
    \Bar{S}_{operator} = \frac{iN}{2\pi} \sum_{\tau-links;\hat{\tau}=0} \Tilde{\varphi}(2\pi n - (\Delta_x^2 + \Delta_y^2)\phi) + iN\sum_{\tau-links;\hat{\tau}=0} \phi \Tilde{n}
\end{equation}
so that fusing the operators and gauging away $n$ and $\Tilde{n}$ yields
\begin{equation}
    S_{combined} = \frac{iN}{2\pi}\sum_{\tau-links;\hat{\tau}=0} \varphi(\Delta_x^2 + \Delta_y^2)(\phi_+-\phi_-), 
\end{equation}
which we interpret as a condensation operator as above.
\subsection{Symmetry Defect}
We consider a symmetry defect at $x=0$.  In this case, the boundary term is 
\begin{equation}
    S_\partial = -iN \sum_{x\tau-squares;\hat{x}=0} \Tilde{\varphi} \Delta_x n_\tau.  
\end{equation}
After summing by parts, we obtain the defect action:
\begin{equation}
    S_{defect} = 2\frac{iN}{2\pi}\sum_{x-links;\hat{x}=0} \phi[-2\pi \Delta_x \Tilde{n}_\tau + \Delta_x\Delta_\tau \Tilde{\varphi}]-iN \sum_{x\tau-squares;\hat{x}=0} \Tilde{\varphi} \Delta_x n_\tau.
\end{equation}
Let us now consider fusing the defect and its opposite.  Repeating the above procedure for the dual theory on $x \leq 0$ gives
\begin{equation}
    \Bar{S}_{defect} = -2\frac{iN}{2\pi}\sum_{x\tau-squares;\hat{x}=0} \Tilde{\varphi}(-2\pi \Delta_x n_\tau + \Delta_x\Delta_\tau \phi) + iN \sum_{x-links;\hat{x}=0} \phi \Delta_x \Tilde{n}_\tau
\end{equation}
so that fusing the defects and gauging away $n_\tau$ and $\Tilde{n}_\tau$ yields
\begin{equation}
    S_{combined} = -2\frac{iN}{2\pi}\sum_{x\tau-squares;\hat{x}=0} \varphi \Delta_x\Delta_\tau (\phi_R-\phi_L),
\end{equation}
which we interpret as a condensation defect as above, with the same caveat as the defect in section 5.

Since the $(n_\tau, n)$ is flat and we are imposing Dirichlet boundary conditions, we can deform the defect, so it is a symmetry.
\section{Discussion and Outlook}
In this paper, we showed that the ``half gauging" construction of non-invertible symmetries applies in field theories with exotic symmetries.  These are all constructed analogously to the duality defect in the $c=1$ compact boson theory.  We gauged a $\mathbb{Z}_N$ subgroup of the $U(1)$ momentum symmetry and employ T-Duality to create an interface.  We did this for the XY Plaquette Model, the XY Cube Model, the 1+1 D Compact Lifshitz Theory, and the 2+1 Dimensional Laplacian Lifshitz theory.  Since our theories are non-relativistic, operators and defects are not generally the same, and we treated them on a case by case basis.  We found noninvertible interfaces that fuse to a condensation defect in all cases.  Our results demonstrate the ability of the methods developed in \cite{NonInvDual,NonInvKW} to peculiar non-relativistic theories.  

Many possibilities for further research remain.  Some worth pursuing are:
\begin{itemize}
    \item Recent work \cite{NIEm} showed that the duality defect in the Ising model is an emanant \cite{Emanant} symmetry - an exact symmetry present in the IR but not in the UV.  In particular, they are the IR avatar of lattice translation.  It would be interesting to examine the analogs symmetries discussed herein on the lattice and see if similar phenomena occur.
    \item Because of the peculiar relationships between the Villain models of Lifshitz theories and their continuum limits, we have left a discussion of the duality symmetry in the continuum for future work.
    \item Most examples of non-invertible symmetries are in relativistic quantum field theory.  It would be interesting to see further examples of non-invertible symmetries in non-relativistic quantum field theories.
    \item Anomalies of these symmetries deserve study.  A promising framework for this is the symmetry TFT \cite{OrbGrpd,TopoSym,CatSymPT,AlgSymCatSym,SymTFTString,SymTFTNI,SymTFTCont,GaugeU1} \footnote{For a recent, similar approach developed for continuous symmetries, see \cite{SymTh}.}, which was recently generalized to include subsystem symmetries in \cite{SubsystemSymTFT}.  
    \item It would be interesting to study the outcome of gauging the exotic momentum symmetries of the Laplacian theories.  
\end{itemize}
\appendix

\acknowledgments

I am grateful for many topical lecture series at TASI 2023: Aspects of Symmetry, especially (for the purposes of this work) those by Sakura Schafer-Nameki, Yifan Wang, Thomas Dumitrescu, Ibou Bah, and Shu-Heng Shao.

We thank Andreas Karch, Po-Shen Hsin, and Shu-Heng Shao for enlightening conversations.

This work was supported by the U.S.~Department of Energy under
Grant DE-SC0022021 and by a grant from the Simons Foundation (Grant
651440, AK).

\appendix
\section{Deforming Interfaces}
In this appendix, we provide a detailed demonstration of the ability to deform the interfaces we discuss in the body of the paper.  We follow the strategy in \cite{NonInvDual}.  In particular, we reduce an initial interface to the deformed interface by integrating out the Lagrange multiplier field in the BF theory, but only on the part of spacetime affected by the deformation.  For specificity, we work with the Modified Villain formulation of the interfaces.  

The key input is the normalization of the BF theory.  One a three dimensional spacetime $X$, for concreteness, the desired partition function is  
\begin{equation}
    \mathcal{Z}[X] = \frac{\vert H^1(X;\mathbb{Z}_N) \vert}{\vert H^0 (X;\mathbb{Z}_N)\vert}.
\end{equation}
The orders of the cohomology groups are simply the number of large gauge transformations.  The partition function counts them.  In order to achieve the desired partition function, the appropriate normalization for the Modified Villain model on an arbitrary triangulation of $X$ is 
\begin{equation}
    \mathcal{Z}[X] = \frac{1}{N^{[1]}}\frac{1}{N^{[0]}}\sum_{a,b}\exp[\frac{2\pi i}{N}\sum_{2-cells}b^{(1)}\Delta a^{(1)}].
\end{equation}
Above, $[k]$ is the number of k-cells.  The normalization counteracts factors of $N$ from fixing the gauge group as one integrates out fields.  

\subsection{Deforming the Defect in the XY Plaquette Model}
We now demonstrate the ability to deform duality defects in the XY Plaquette model.  The normalization for the exotic $\mathbb{Z}_N$ gauge theory that accounts for factors of $N$ is
\begin{equation}
    \mathcal{Z} = \frac{1}{N^{[c]}}\frac{1}{N^{[v]}}\sum_{\varphi^{xy},\hat{n}_\tau,\hat{n}_{xy}} exp[\frac{2\pi i}{N}\sum_{cubes}\varphi^{xy}(\Delta_\tau \hat{n}_{xy} - \Delta_x \Delta_y \hat{n}_\tau)]. 
\end{equation}
Above, $[c]$ is the number of cubes and $[v]$ is the number of vertices.  Now, consider a defect at a fixed $y$.  We seek to deform it by a cube.  If the partition function of the system with the cube added is the same as the partition function without the cube, we can deform the defect.  Taking into account the Dirichlet boundary conditions on the defect, we note that the $\mathbb{Z}_N$ gauge theory with an extra cube has an additional cube and four additional vertices.  We integrate out $\varphi^{xy}$ on the cube, generating a factor of $N$ and forcing $\Delta_\tau \hat{n}_{xy} - \Delta_x \Delta_y \hat{n}_\tau=0$.  We can gauge away the two $\hat{n}_{xy}$ and the two $\hat{n}_\tau$ on the cube.  This uses four gauge parameters, generating a factor of $N^4$.  This cancels the factors of $N$ in the partition function with an additional cube, showing that we can deform the defect.  An identical argument can be made switching the roles of $x$ and $y$.  Note that the ability to deform the defect is guaranteed by the normalization of the $\mathbb{Z}_N$ gauge theory and the flatness of $(\hat{n}_\tau,\hat{n}_{xy})$.
\subsection{Deforming the Defect in the Laplace-Lifshitz Model}
We now demonstrate the ability to deform duality defects in the Laplacian Lifshitz theory.  The normalization of the Laplacian $\mathbb{Z}_N$ gauge theory that accounts for factors of $N$ is 
\begin{equation}
    \mathcal{Z} = \frac{1}{N^{[\tau]}}\frac{1}{N^{[v]}}\sum_{\Tilde{m},\hat{n}_{xx},\hat{n}_\tau} \exp[\frac{2\pi i}{N}\sum_\tau \Tilde{m}(\Delta_\tau \hat{n}_{xx} - \Delta_x^2 \hat{n}_\tau)].
\end{equation}
Above, $[\tau]$ is the number of $\tau$ links and $[v]$ is the number of vertices.  Now, consider a defect at a constant $x$.  We seek to deform it by a square.  As before, we can deform the defect if the partition function is unchanged by adding a square.  Taking into account the Dirichlet boundary conditions on the defect, the $\mathbb{Z}_N$ gauge theory with an extra square contains an extra $\tau$ link and two additional vertices.  We integrate out $\Tilde{m}$ on the $\tau$ link, generating a factor of $N$ and forcing $\Delta_\tau \hat{n}_{xx} - \Delta_x^2 \hat{n}_\tau = 0$.  We can gauge away the $\hat{n}_\tau$ and two $\hat{n}_{xx}$.  Thus uses two gauge parameters, generating a factor of $N^2$.  The net $N^3$ cancels the change in normalization from the extra square, showing that we can deform the defect.  Note that the ability to deform the defect is guaranteed by the normalization of the $\mathbb{Z}_N$ gauge theory and the flatness of $(\hat{n}_\tau,\hat{n}_{xx})$.




\end{document}